\documentclass{article} 
\usepackage{graphicx}
\usepackage[colorlinks=true, allcolors=blue,breaklinks]{hyperref}
\usepackage{float}
\usepackage{xurl}
\usepackage{blindtext}
\usepackage{setspace}
\usepackage{subfig}
\title{EEG Synthetic Data Generation Using Probabilistic Diffusion Models}

\author{Giulio Tosato$^1$\\  \href{mailto:g.tosato@tilburguniveristy.edu}{g.tosato@tilburguniversity.edu} \and Cesare M. Dalbagno$^1$\\  \href{mailto:c.m.dalbagno@tilburguniversity.edu}{c.m.dalbagno@tilburguniversity.edu} \and Francesco Fumagalli\ $^1$\\  \href{mailto:f.fumagalli@tilburguniversity.edu}{f.fumagalli@tilburguniversity.edu}}

\date{February 2023}
\begin{document} 
\maketitle
\begin{abstract}
Electroencephalography (EEG) plays a significant role in the Brain Computer Interface (BCI) domain, due to its non\textendash invasive nature, low cost, and ease of use, making it a highly desirable option for widespread adoption by the general public. This technology is commonly used in conjunction with deep learning techniques, the success of which is largely dependent on the quality and quantity of data used for training. To address the challenge of obtaining sufficient EEG data from individual participants while minimizing user effort and maintaining accuracy, this study proposes an advanced methodology for data augmentation: generating synthetic EEG data using denoising diffusion probabilistic models. The synthetic data are generated from electrode-frequency distribution maps (EFDMs) of emotionally labeled EEG recordings. To assess the validity of the synthetic data generated, both a qualitative and a quantitative comparison with real EEG data were successfully conducted. This study opens up the possibility for an open\textendash source accessible and versatile toolbox that can process and generate data in both time and frequency dimensions, regardless of the number of channels involved. Finally, the proposed methodology has potential implications for the broader field of neuroscience research by enabling the creation of large, publicly available synthetic EEG datasets without privacy concerns.
\end{abstract}
\newpage
\section{Introduction}
The use of Electroencephalogram (EEG) recordings for Brain-Computer Interface (BCI) has gained significant attention in the scientific community due to its various applications (A. Roman-Gonzalez, 2012). EEG-BCI technology has been used to develop communication tools  as well as rehabilitation and assistive technologies
for disabled individuals, such as those
in a complete locked-in state or with severe motor impairments (Minguillon et al., 2017; Neuper et al., 2006). In the future, EEG-based BCIs have the potential to enhance human performance, connecting prosthetics and remotely controlled devices to allow for greater physical and mental abilities (Bright et al., 2016; Christoph Guger \& Carin Hertnaes, n.d.; Penghai \& Baikun, 2007; Scherer et al., 2004).\\

However, the practical implementation of EEG-BCI technologies is hindered by limitations such as low quality and availability of EEG data, difficulty in collecting long recordings, low spatial resolution, and the need for a large amount of data to train accurate models (Mona Sazgar \& Michael G. Young, 2019).\\

The proposed solution is to use a denoising diffusion probabilistic model (DDPM) for EEG data synthesis, which was proven to outperform Generative Adversarial Networks (GANs, Yun Luo, 2018; Shovon et al., 2019) in image generation (P. Dhariwal, 2021). Even though a similar study, parallel to ours, has just been conducted (Torma \& Szegletes, 2023), an important question remained still unanswered: is the model producing novel data or simply replicating examples from the original dataset?\\

The focus of our work is to address this question while also creating an open-source synthetic EEG toolbox freely available and easily accessible for specific usage. 
The diffusion model was trained on electrode-frequency distribution maps (EFDMs, Wang et al., 2020) developed from a large emotion-labeled EEG dataset, due to the fact that frequency features assure better accuracy in classifying EEG signals (Huang et al., 2022; J. Wang et al., 2017; X.-W. Wang et al., 2011). To demonstrate the quality and novelty of the generated data, we compared the accuracy of a classifier on both original and synthetic data.

\section{Methods}
\subsection{EEG}
Electroencephalography (EEG) is a widely used non-invasive brain imaging technique that measures the electrical activity of the brain using electrodes placed on the scalp (Niedermeyer and da Silva, 2005). EEG recordings are made by recording the voltage fluctuations between electrodes, which reflect the electrical activity of the brain. EEG has high temporal resolution, low cost, and great portability, making it ideal for studying brain function in the fields of neuroscience, psychology, and clinical medicine (Darvas et al., 2004). Additionally, compared to other brain imaging techniques, it is safer, because no ionizing radiation nor harmful substances are used during the recording process (Prasad \& Tvk, 2014; M. Teplan, 2002). EEG is particularly useful for studying oscillatory brain activity, rhythmic patterns of neural activity that occur at specific frequencies. These oscillations are thought to play a role in cognitive and perceptual processes such as perception, attention, and memory (Tallon-Baudry and Bertrand, 1999).

\subsection{Short-time Fourier Transform}
The Short-time Fourier Transform (STFT) is a widely used technique in analysing the frequency properties of time series data. It operates by dividing a nonstationary time series into smaller segments and performing a discrete Fourier transform (DFT) on each segment. Unlike the fast Fourier transform (FFT), the STFT provides time-localized frequency information, making it well suited for signals that exhibit changes in frequency content or contain transient events (Shovon et al., 2019; F. Wang et al., 2020).

The FFT, a popular method for performing the DFT and analyzing signals, assumes that the time series being analyzed are stationary. However, this is often not the case for signals such as EEG recordings (Paluš, 1996). The STFT overcomes this limitation by computing the DFT over windowed segments of the signal, making it a powerful tool for analyzing nonstationary signals.

\subsection{Denoising probabilistic diffusion models}
\begin{figure}[h]
\centering
\includegraphics[width=1\textwidth]{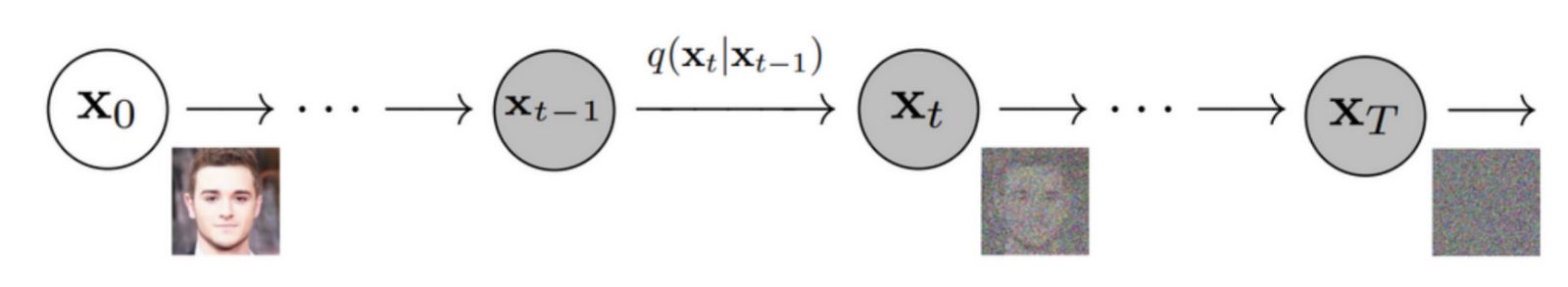}
\caption{\label{fig:noise} Progressive addition of Gaussian noise.}
\end{figure}

DDPMS are a type of generative model that uses a parameterized Markov chain and variational inference to generate synthetic samples that match real data (Ho et al., 2020). The transitions of this chain are trained to reverse the diffusion process, which on its own is a gradual introduction of Gaussian noise in the data. The diffusion process is represented as a latent variable model that uses a fixed Markov chain to map to the latent space. The purpose of training a diffusion model is to learn how to perform the reverse process, which can be used to generate synthetic data (Introduction to Diffusion Models for Machine Learning, 2022). The chain gradually introduces noise to the data until the signal is fully obliterated, and only noise remains (Figure \ref{fig:noise}).

\begin{figure}[h]
\centering
\includegraphics[width=1\textwidth]{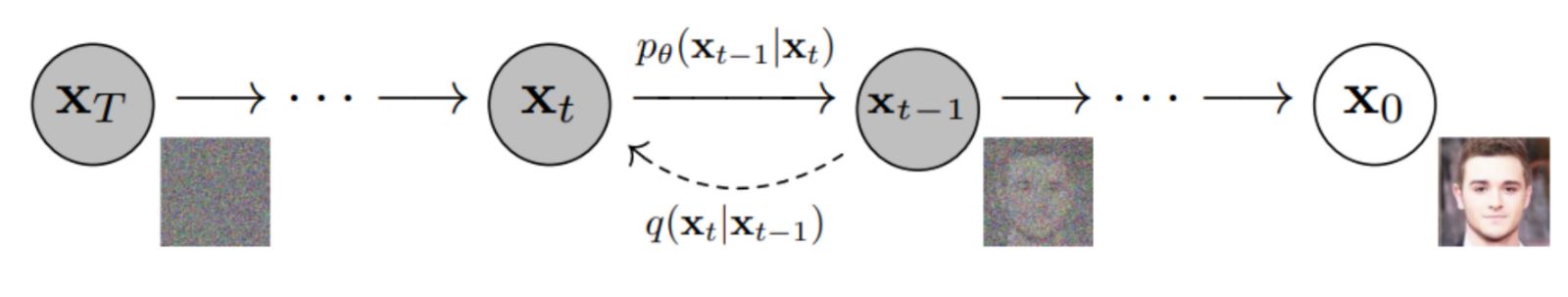}
\caption{\label{fig:denoise} Progressive subtraction of Gaussian noise.}
\end{figure}

The approximate posterior distribution q(x1:T|x0) (Figure \ref{fig:denoise}), is obtained, where x1, ...xT are the latent variables with the same dimensionality as x0. The final asymptotic transformation of the data is to pure Gaussian noise.

\subsection{OpenAI improved-diffusion model}
The \href{https://github.com/openai/improved-diffusion} { OpenAI improved-diffusion} model is a cutting-edge deep learning model that has demonstrated exceptional performance on a diverse range of datasets (Nichol \& Dhariwal, 2021), as evidenced by numerous citations in the field. We aim to create a versatile toolbox that can handle a wide variety of tasks, and the improved-diffusion model represents the best compromise between optimized performance and optimized code. This allows us to create a powerful tool that is accessible to a wide range of users, including researchers outside the field of AI.

\subsection{Classifier}
\begin{figure}[h]
\hspace*{-1cm} 
\centering
\includegraphics[width=1.2\textwidth]{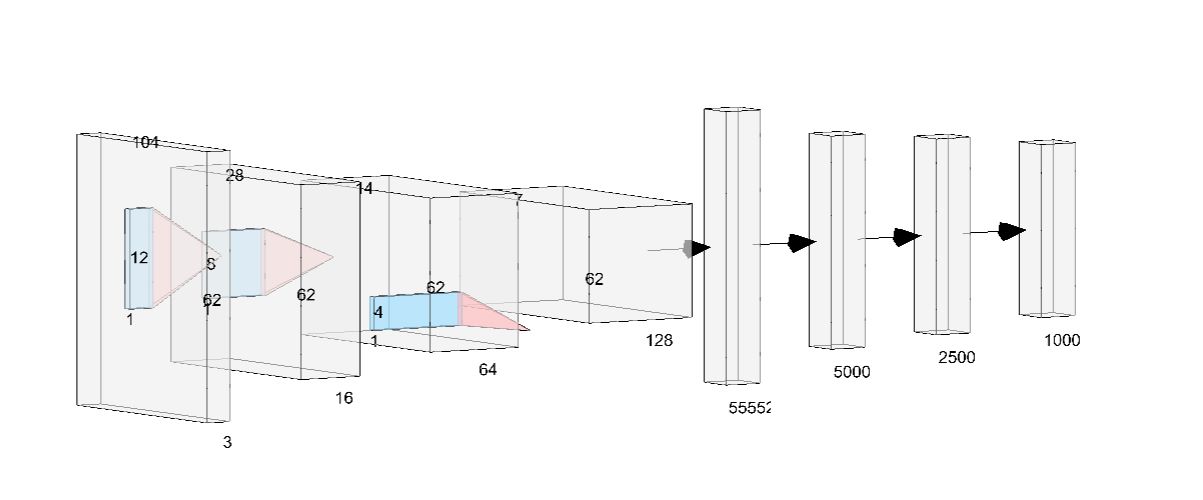}
\caption{\label{fig:classifier_architecture} Classifier architecture.}
\end{figure}

In this study, a classifier was developed using the Pytorch framework (Paszke et al., 2017). Initially, the model was trained with two classes, but it was designed with the capability to handle multiple classes to enhance its versatility. Thus, CrossEntropyLoss was adopted as loss function. The model was optimized with a batch size of 128, a learning rate of 0.0001, and the GELU activation function. The specific details of the model configuration can be found in Table \ref{tab:classifier_architecture} and illustrated in Figure \ref{fig:classifier_architecture}.

\begin{table}[h]
\centering
\begin{tabular}{|c|c|c|}
\hline
\textbf{Layer Type} & \textbf{Input Shape} & \textbf{Hyperparameters} \\ \hline
Conv2d & 3x104x62 & kernel\_size=(12, 1), stride=(1, 1), padding=(2, 0) \\ \hline
MaxPool2d & 16x28x62 & kernel\_size=(4, 1), stride=(4, 1), padding=0, \\ \hline
Conv2d & 3x14x62 & kernel\_size=(8, 1), stride=(1, 1), padding=(1, 0) \\ \hline
MaxPool2d & 64x14x & kernel\_size=(2, 1), stride=(2, 1), padding=0,  \\ \hline
Conv2d & 3x7x62 & kernel\_size=(4, 1), stride=(1, 1), padding=(1, 0) \\ \hline
MaxPool2d & 128x100x62 & kernel\_size=(2, 1), stride=(2, 1), padding=0 \\ \hline
LazyLinear & 3x100x62 & in\_features=0, out\_features=5000, bias=True \\ \hline
Linear & 5000 & in\_features=5000, out\_features=2500, bias=True \\ \hline
Linear & 2500 & in\_features=2500, out\_features=1000, bias=True \\ \hline
Linear & 1000 & in\_features=1000, out\_features=2, bias=True \\ \hline
\end{tabular}
\caption{\label{tab:classifier_architecture} Classifier architecture.}
\end{table}

\subsection{Hardware used}
Intel(R) Xeon(R) Gold 6346 CPU @ 3.10GHz equipped with 2 NVIDIA A40 45 GB of ram.

\section{Procedure}
For this study, the \href{https://bcmi.sjtu.edu.cn/home/seed/} { SEED V} dataset provided by Shanghai Jiao Tong University was used. Specifically, the EEG raw labeled data from 16 participants in three sessions. During each session, the participants were presented with four video clips designed to elicit emotions such as Disgust, Fear, Neutral, Sadness, or Happiness. Only happy and sad data were used in our study.

\subsection{EFDMs creation}

The preprocessing of the EEG data and the creation of Electrode-frequency Distribution Maps (EFDMs, Wang et al., 2020) were conducted according to the pipeline available on the \href{https://github.com/DevJake/EEG-diffusion-pytorch} {GitHub repository}. The MNE-Python library (Gramfort et al., 2013) was used to read the raw data and create a two-dimensional matrix, excluding the unlabeled time steps, with channels as columns and time steps as rows. The Short-time Fourier Transform (STFT) was then applied to the data using the mne.time\_frequency.stft function. The modules of the real and imaginary part of each value in the matrix were calculated and normalized cutting up to 100Hz, resulting in 2D arrays known as the EFDMs. These are grayscale images that plot the intensity values (µV) in each channel (1-62) at each frequency (up to 100Hz) each transform step.
The EFDMs were first squared (128x128) by adding padding, flipped upside down and then converted to RGB images by multiplying by 255 and converting to np.uint8 to meet the input format the model required. Two pickle lists, one for each emotion, were created, containing the EFDMs for each of the 3 sessions for each of the 16 participants.

\subsection{Classifier training}
The classifier was trained using the same dataset as the diffusion model: 24000 images per emotion; its accuracy was assessed using a separate dataset of 6000 images per emotion previously unseen by both the classifier and the diffusion model. 

\subsection{Diffusion model training}
\begin{figure}[H]
\centering
\begin{minipage}{0.25\textwidth}
\centering
\includegraphics[width=1\linewidth]{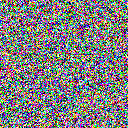}
\caption*{\small(a) 0 epochs}
\end{minipage}%
\hspace{0.05\textwidth}
\begin{minipage}{0.25\textwidth}
\centering
\includegraphics[width=1\linewidth]{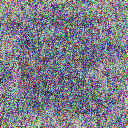}
\caption*{\small(b) 10 epochs}
\end{minipage}%
\hspace{0.05\textwidth}
\begin{minipage}{0.25\textwidth}
\centering
\includegraphics[width=1\linewidth]{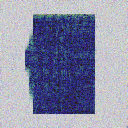}
\caption*{\small(c) 20 epochs}
\end{minipage}
\vspace{1cm}

\begin{minipage}{0.25\textwidth}
\centering
\includegraphics[width=1\linewidth]{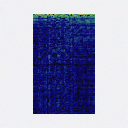}
\caption*{\small(d) 30 epochs}
\end{minipage}%
\hspace{0.05\textwidth}
\begin{minipage}{0.25\textwidth}
\centering
\includegraphics[width=1\linewidth]{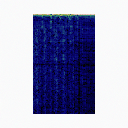}
\caption*{\small(e) 40 epochs}
\end{minipage}%
\hspace{0.05\textwidth}
\begin{minipage}{0.25\textwidth}
\centering
\includegraphics[width=1\linewidth]{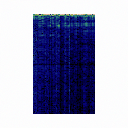}
\caption*{\small(f) 50 epochs}
\end{minipage}

  \caption{\label{tab:Image_generated}Images generated by the Diffusion Model during training (each image is generated upside down, with the x-axis representing the channels and the y-axis representing the frequencies).}
\end{figure}

We adapted the OpenAI improved diffusion model to our needs (code available at our \href{https://github.com/DevJake/EEG-diffusion-pytorch} {GitHub repository}). \\
To train the diffusion model (Figure \ref{tab:Image_generated}), the following parameters were used: \\
MODEL\_FLAGS="\--image\_size 128 --num\_channels 128 --num\_res\_blocks 3"\\
DIFFUSION\_FLAGS="--diffusion\_steps 1000 --noise\_schedule linear"\\
TRAIN\_FLAGS="--lr 1e-4 --batch\_size 32"\\

\section{Results}

\begin{figure}[H]
\centering
\includegraphics[width=1\textwidth]{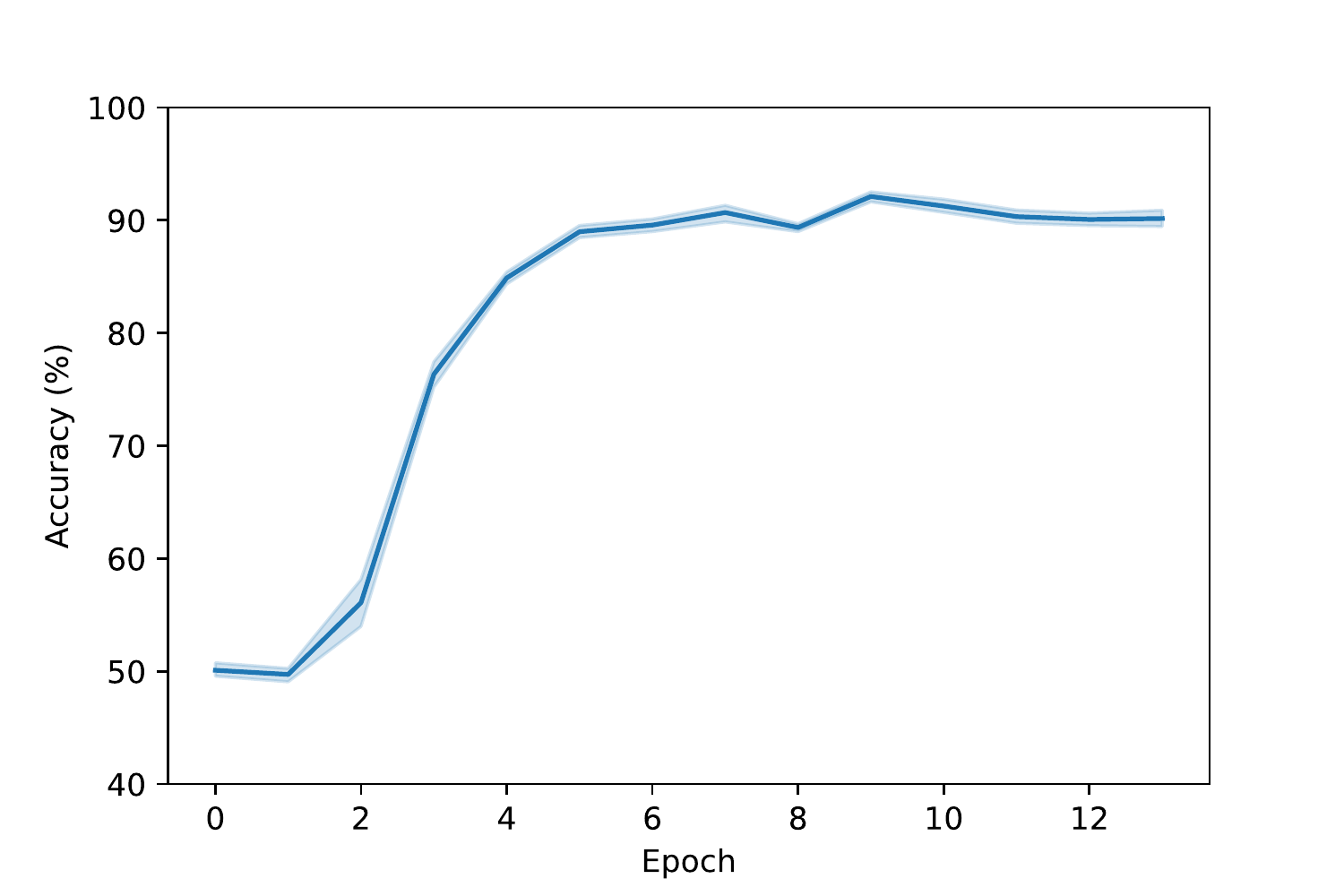}
\caption{\label{fig:classifier_on_diffusion} Evaluation accuracy of the classifier trained 20 times on original data with different initializations and evaluated on synthetic data generated by the diffusion model at different epochs represented in the x-axis (the solid line represents the average, the shadow represents the 95\% confidence interval).}
\end{figure}

We tested the performance of the classifier trained on real data for 20 epochs on synthetic data to assess whether the generated data were reliable and if they could improve the performance of the classifier model (Figure~\ref{fig:classifier_on_diffusion}).

Initially, there were concerns that the diffusion model may have overfitted the training data, resulting in the generation of mere copies of the original dataset rather than creating novel samples (Somepalli et al., 2022). To test this, the classifier was trained on both synthetic and real data and then tested on never-before-seen real data. The performance of this model was compared with that of a model trained solely on real data, under the assumption that if generated data did not provide additional information, the performances of the two would be identical.

\begin{figure}[H]
\begin{minipage}{0.5\textwidth}
\centering
\includegraphics[width=1\textwidth]{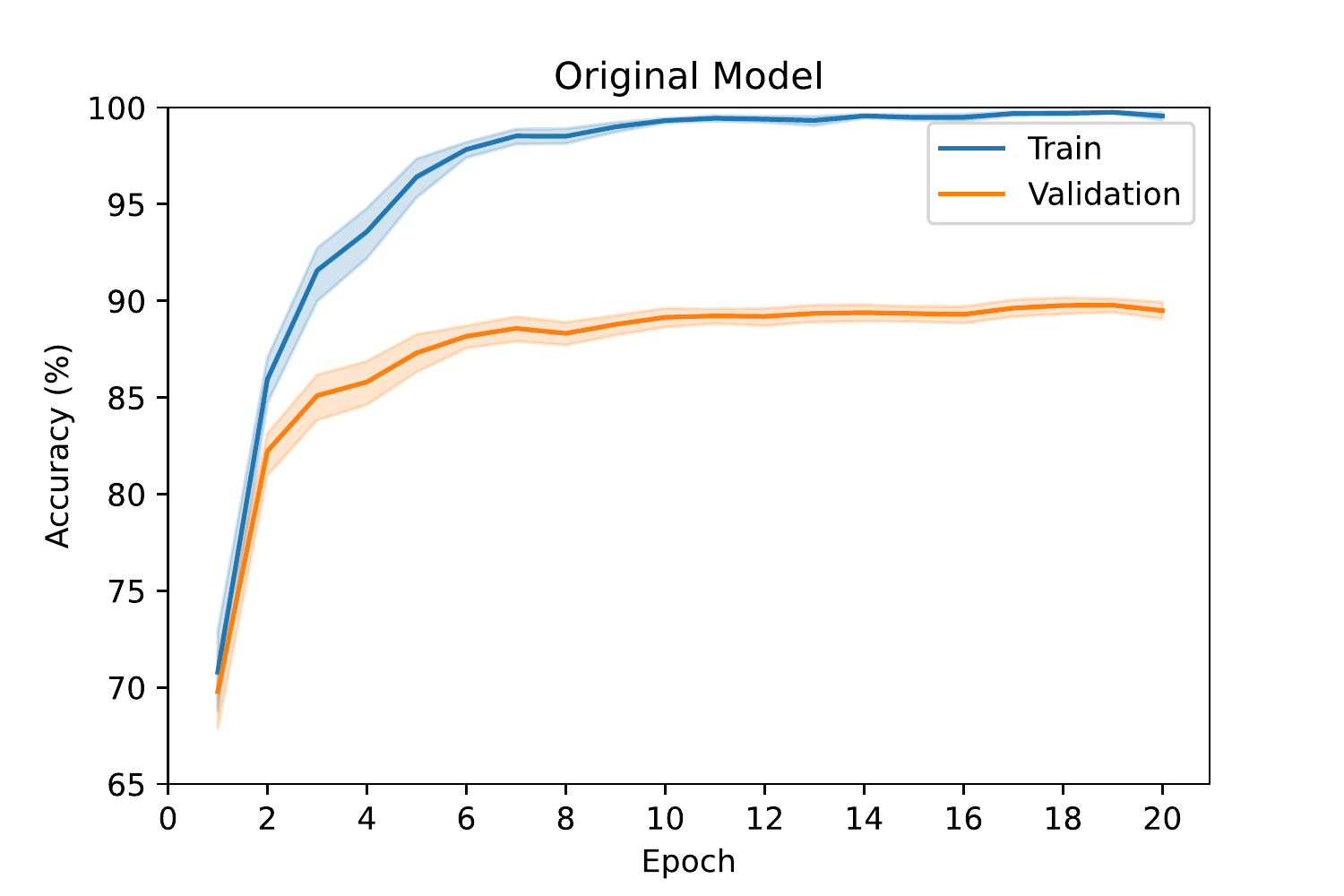} 
\caption*{\small(a)}
\end{minipage}%
\hspace{0.05\textwidth}
\begin{minipage}{0.5\textwidth}
\centering
    \includegraphics[width=1\textwidth]{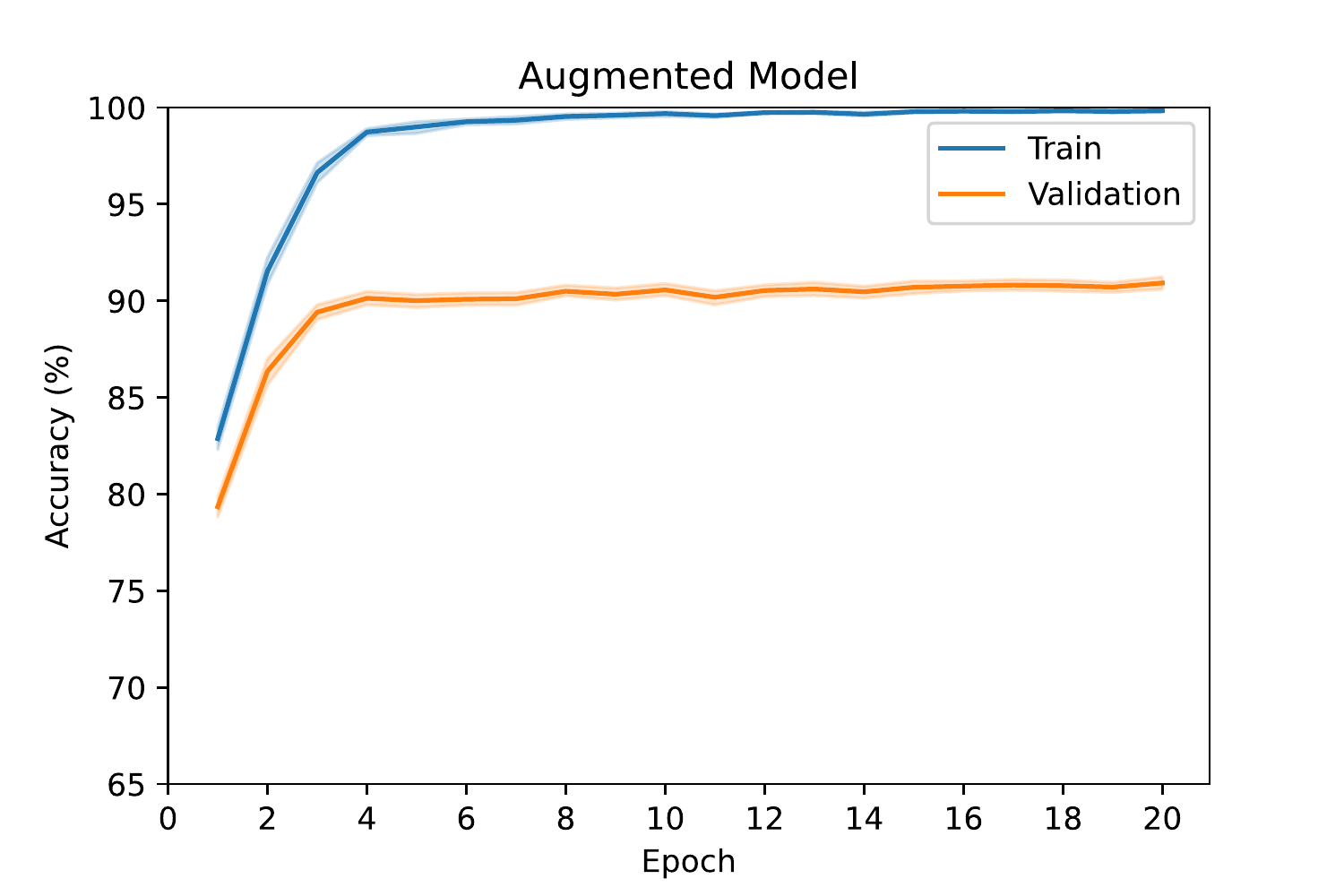}
\caption*{\small(b)}
\end{minipage}
  \caption{\label{fig:classifier_accuracy}Comparison of train and validation accuracy averaged over 20 runs for both models (the solid line represents the average, the shadow represents the 95\% confidence interval).}
\end{figure}

\begin{figure}[H]
\begin{minipage}{0.5\textwidth}
\centering
\includegraphics[width=1\textwidth]{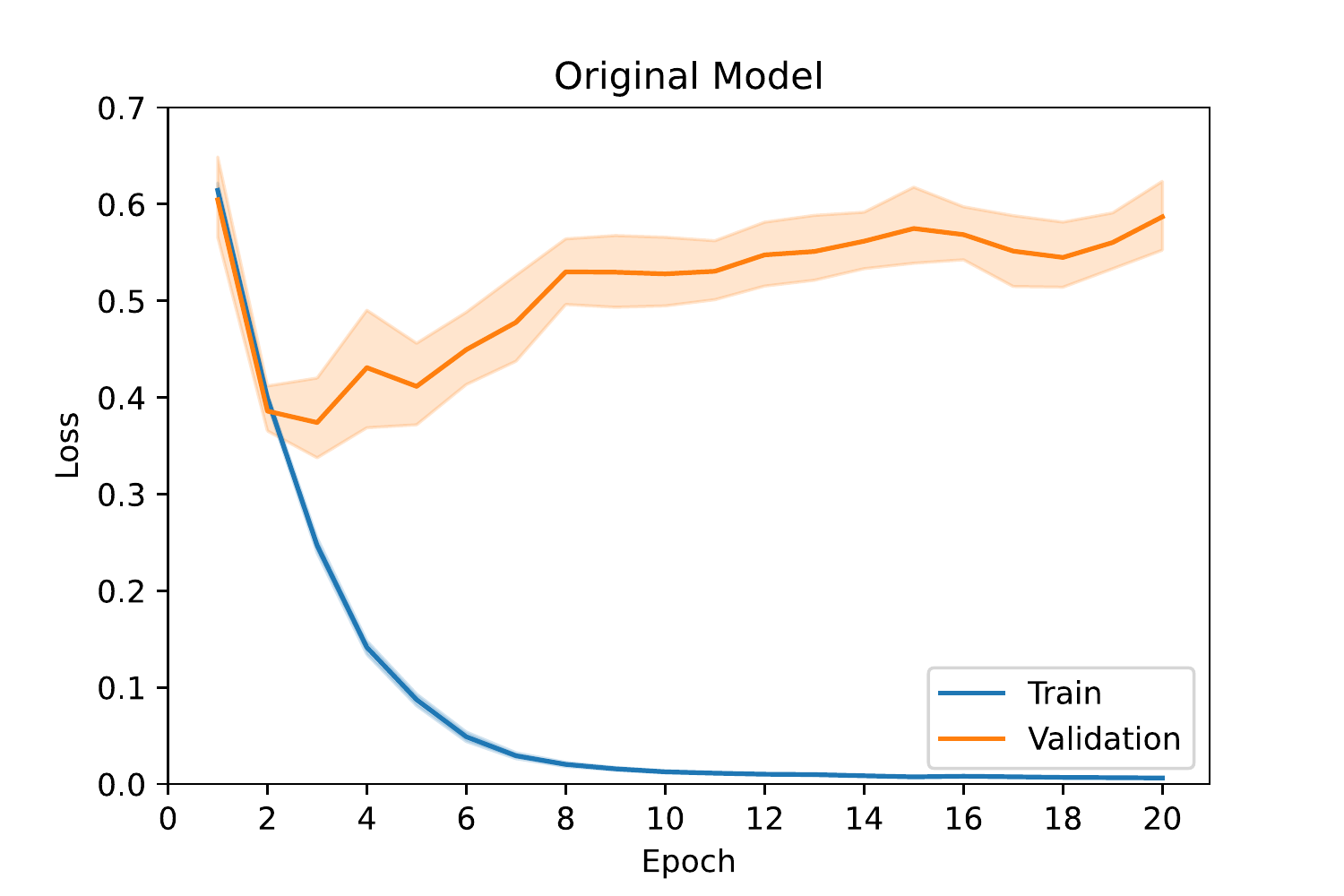} 
\caption*{\small(a)}
\end{minipage}%
\hspace{0.05\textwidth}
\begin{minipage}{0.5\textwidth}
\centering
\includegraphics[width=1\textwidth]{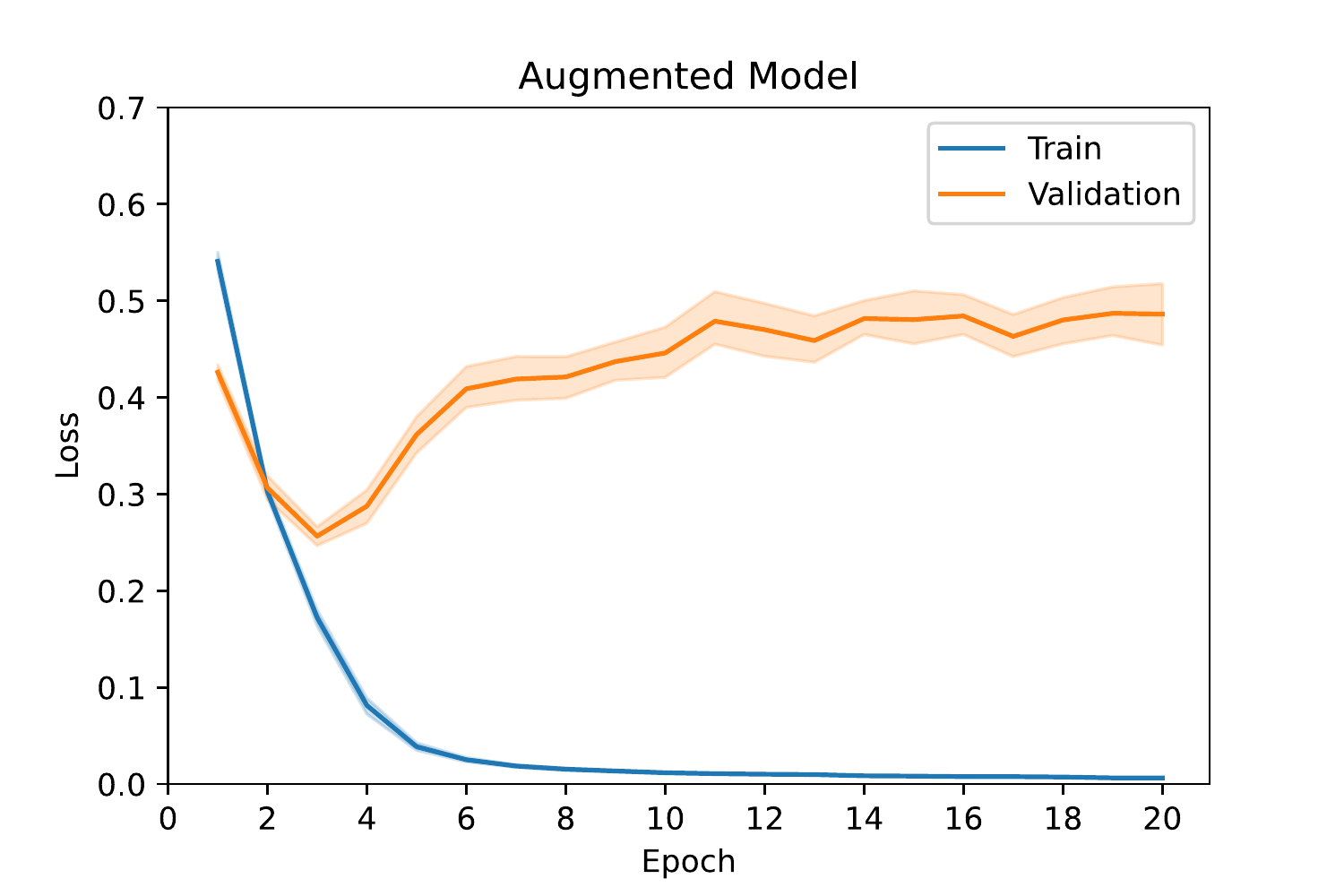}
\caption*{\small(b)}
\end{minipage}
\caption{\label{fig:classifier_loss}Comparison of train and validation loss averaged over 20 runs for both models (the solid line represents the average, the shadow represents the 95\% confidence interval).}
\end{figure}

To mitigate the impact of chance on the classifier's performance, a two-fold training approach was employed. The classifier was first trained 20 times with different random states using only real images (trained on 24000 and tested on 6000 images per emotion), followed by 20 more training with a combination of real (24000 per emotion) and synthetic images (15000 per emotion) then tested on 6000 images per emotion never seen before by both models (Figure~\ref{fig:classifier_accuracy} and Figure~\ref{fig:classifier_loss}). This method allowed for a more robust and reliable evaluation of the classifier's accuracy (Torch.Manual\_seed(3407) Is All You Need, 2021).

\begin{figure}[H]
\begin{minipage}{1\textwidth}
\caption*{\small(a)}
\vspace*{-4mm}
\centering
\includegraphics[width=1\textwidth]{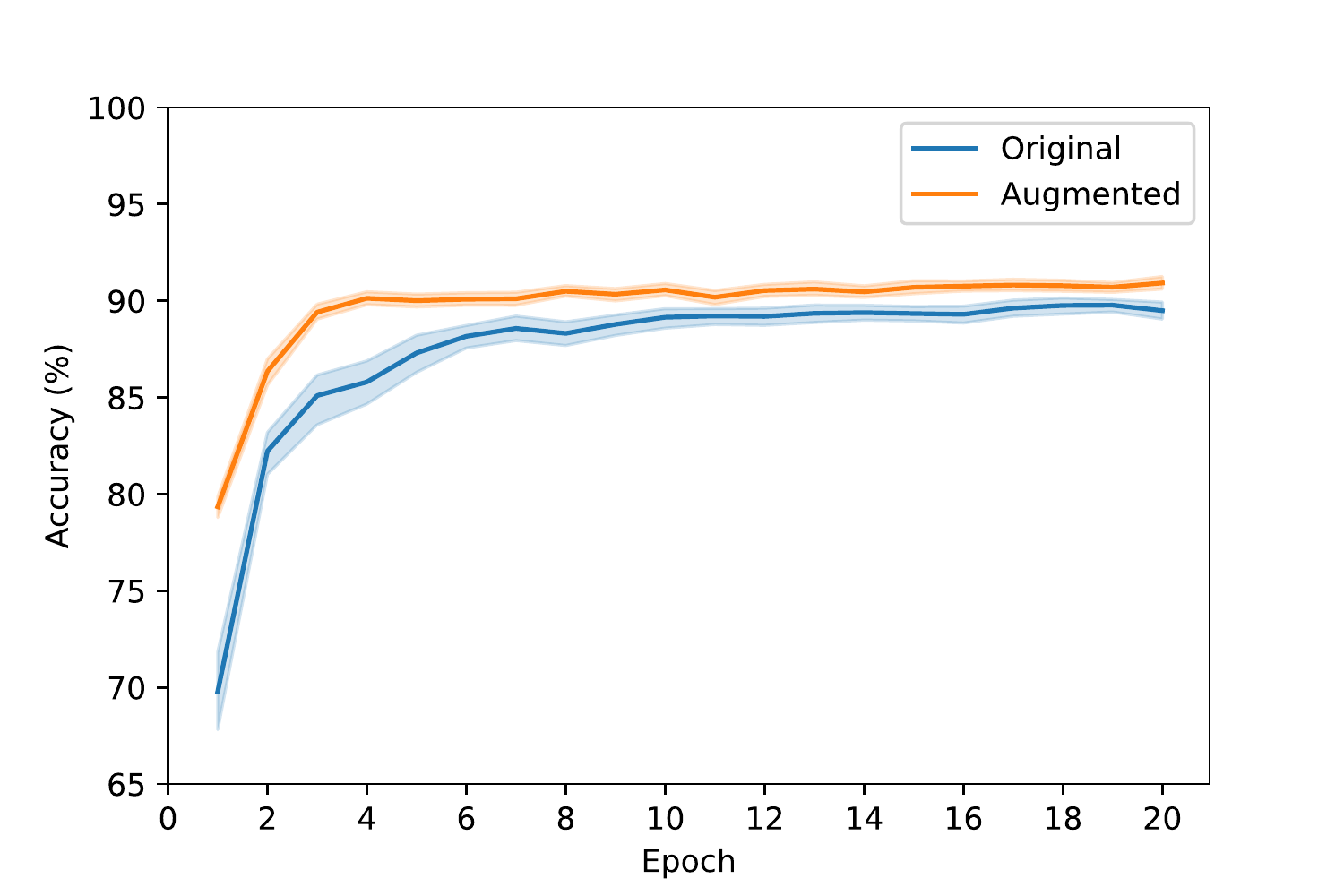}
\captionsetup{justification=raggedright,singlelinecheck=false}
\vspace*{3mm}
\end{minipage}%
\hspace{0.05\textwidth}
\begin{minipage}{1\textwidth}
\caption*{\small(b)}
\vspace*{-4mm}
\centering
\includegraphics[width=1\textwidth]{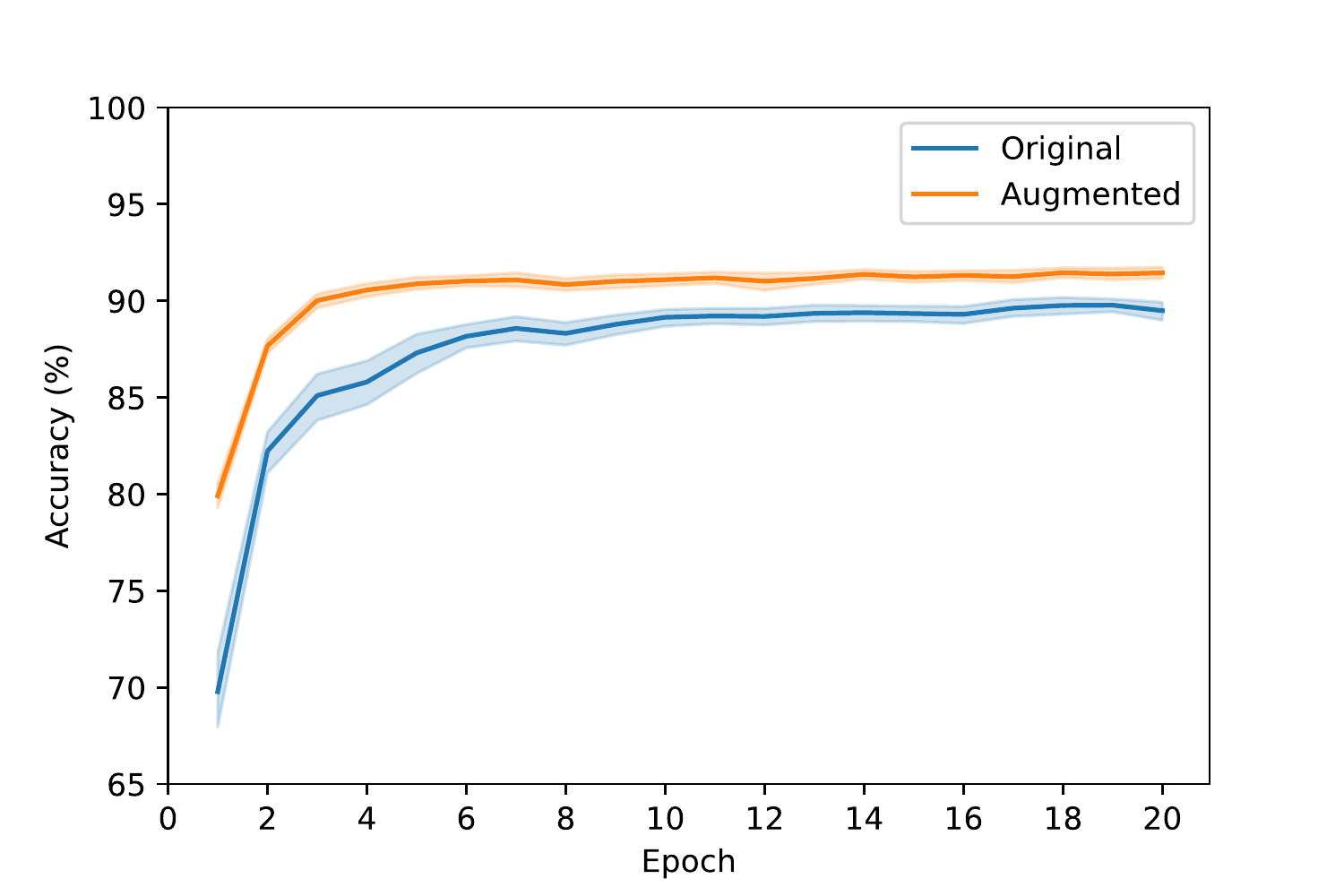}
\captionsetup{justification=raggedright,singlelinecheck=false}
\end{minipage}
\caption{\label{fig:classifier_accuracy_comparison}Comparison of validation accuracy between 20 runs of the same model trained with original data only ("Original") and 20 runs trained with both original and synthetic data ("Augmented"). Synthetic data were generated by a diffusion model trained for 40 epochs (a) and 60 epochs (b) (the solid line represents the average, the shadow represents the 95\% confidence interval).}
\end{figure}

The improvement observed in Figure~\ref{fig:classifier_accuracy_comparison} and Table \ref{tab:table_comparison_acc}, suggests that the model is not just replicating the original data, but creating synthetic data that can be useful for data augmentation. The results of this study highlight the potential of using synthetic data in combination with real data to improve the accuracy of classifiers.

\begin{table}[h!]
\centering
\begin{tabular}{|c|c|}
\hline
\textbf{Classifier type} & \textbf{Max. Average Accuracy}  \\ \hline
Original & 91.434  \\ \hline
Augmented 40 epochs & 92.634  \\ \hline
Augmented 60 epochs & 92.984  \\ \hline
\end{tabular}
\caption{\label{tab:table_comparison_acc}Accuracy comparison between the models trained 20 times with real data only ("Original"), and the models trained 20 times with both original and synthetic data generated by the diffusion model trained for 40 epochs ("Augmented 40 epochs") and 60 epochs ("Augmented 60 epochs").}
\end{table}

\begin{figure}[H]
\begin{minipage}{0.5\textwidth}
\centering
\includegraphics[width=1\linewidth]{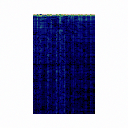}
\vspace*{-12mm}
\caption*{\small(a)}
\end{minipage}%
\hspace{0.05\textwidth}
\begin{minipage}{0.5\textwidth}
\centering
\includegraphics[width=1\linewidth]{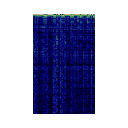}
\vspace*{-12mm}
\caption*{\small(b)}
\end{minipage}
  \caption{\label{fig:final}Comparison between a synthetic image generated by the Diffusion Model trained for 60 epochs (left) and an original image (right).}
\end{figure}

\section{Discussion}

\subsection{Future directions}
In our forthcoming iteration of the paper, we aim to enhance the efficiency, precision, and versatility by directly processing arrays in the diffusion model. This optimization will streamline the pipeline by eliminating the need for converting between arrays and images. Furthermore, the modified model will accept arrays of any 3d shape as inputs; we will also compare the results with data obtained from time series, where the x-axis represents the time domain and the y-axis represents the EEG cap channels. Our experiments have demonstrated that the diffusion model can process images with a resolution up to 256x256 pixels, enabling the processing of approximately 33 seconds of EEG signals from an 8-channel EEG cap with a sampling rate of 250Hz (8x250x33). The capability of the diffusion model to generate ampler time windows in the data can facilitate more complex analyses of brain activity. To further advance the significance of our work, we plan to explore ways to create connected and coherent samples, such as slices of a single long recording.

In addition, we know that brains are different, and this difference results in unique EEG registered activity (Riding et al., 1997; Smit et al., 2012): this poses big challenges in training patient-specific deep learning models due to the large amount of data required. A last idea is to adapt the concept of Diffusion Model few-shot learning (Giannone et al., 2022) to fine-tune a large model pre-trained on multiple people's data using a limited amount of data recorded on a single individual to create a large personalized dataset, therefore improving the performance of the model being used for their specific needs. This technique could cut down sampling time and make it cheaper and more efficient to get an EEG scan, since just a couple of minutes of recording could be augmented into ample datasets. 

\subsection{Possible limitations}
Diffusion models are a powerful tool, yet computationally expensive. In a following update, we will compare the quality of DDPMs' generated data with data generated using more traditional methods, such as introducing random Gaussian noise in the dataset. If the diffusion-generated data perform better on the same classification test, it will be an indication that diffusion models actually are one of the best ways to create accurate synthetic samples, fundamental for tasks involving human brain enhancement.

\section{Conclusion}
Our study aimed to address the issue of limited availability of high-quality and accessible EEG data, which has become increasingly important in various research areas, including rehabilitation and enhancement of human capacities.
In this study, we proposed a solution to this problem by using data augmentation via  DDPMs. Our classifier, trained on real data only, achieved over 90\% average accuracy when tested on our generated samples (Figure \ref{fig:classifier_on_diffusion}), indicating that this type of data can be used as useful samples to augment the original dataset. To validate that the DDPM was not simply replicating the training data, we compared the performance of a classifier trained on both synthetic and real data to that of a model trained solely on real data (Figure \ref{fig:classifier_accuracy_comparison}). The hybrid model consistently outperformed the original classifier, indicating that the generated data contains information that cannot be found in the original data.

This finding provides evidence that generated data can improve the performance of EEG related applications. Furthermore, the generated data can be shared without any privacy concern since they are not directly sampled from individuals. Future versions of this paper will make the toolbox available on our GitHub repository more versatile and accessible, thereby creating a useful instrument for data generation in various fields. 

\newpage

\section{Bibliography}

\doublespacing
Avid Roman-Gonzalez. (2012). EEG Signal Processing for BCI Applications: Vol. 98 (1). HAL. \url{https://hal.science/hal-00742211} 

Bracewell, R. N. (1989). The Fourier Transform. Scientific American, 260(6), 86–95. \url{https://www.jstor.org/stable/24987290}

Bright, D., Nair, A., Salvekar, D., \& Bhisikar, S. (2016). EEG-based brain controlled prosthetic arm. 2016 Conference on Advances in Signal Processing (CASP), 479–483. \url{https://doi.org/10.1109/CASP.2016.7746219}

Christoph Guger, W. H., \& Carin Hertnaes, G. P. (n.d.). Prosthetic Control by an EEG-based BrainComputer Interface (BCI).

Darvas, F., Pantazis, D., Kucukaltun-Yildirim, E., \& Leahy, R. M. (2004). Mapping human brain function with MEG and EEG: Methods and validation. NeuroImage, 23, S289–S299. \url{https://doi.org/10.1016/j.neuroimage.2004.07.014}

Dhariwal, P., \& Nichol, A. Q. (2022, January 26). Diffusion Models Beat GANs on Image Synthesis. Advances in Neural Information Processing Systems. \url{https://openreview.net/forum?id=AAWuCvzaVt}

Giannone, G., Nielsen, D., \& Winther, O. (2022). Few-Shot Diffusion Models (arXiv:2205.15463). arXiv. \url{https://doi.org/10.48550/arXiv.2205.15463}

Gramfort, A., Luessi, M., Larson, E., Engemann, D., Strohmeier, D., Brodbeck, C., Goj, R., Jas, M., Brooks, T., Parkkonen, L., \& Hämäläinen, M. (2013). MEG and EEG data analysis with MNE-Python. Frontiers in Neuroscience, 7. \url{https://www.frontiersin.org/articles/10.3389/fnins.2013.00267}

Ho, J., Jain, A., \& Abbeel, P. (2020). Denoising Diffusion Probabilistic Models. Advances in Neural Information Processing Systems, 33, 6840–6851. \url{https://proceedings.neurips.cc/paper/2020/hash/4c5bcfec8584af0d967f1ab10179ca4b-Abstract.html}

Huang, E., Zheng, X., Fang, Y., \& Zhang, Z. (2022). Classification of Motor Imagery EEG Based on Time-Domain and Frequency-Domain Dual-Stream Convolutional Neural Network. IRBM, 43(2), 107–113. \url{https://doi.org/10.1016/j.irbm.2021.04.004}

Introduction to Diffusion Models for Machine Learning. (2022, May 12). News, Tutorials, AI Research. \url{https://www.assemblyai.com/blog/diffusion-models-for-machine-learning-introduction/}

Luo, Y., \& Lu, B.-L. (2018). EEG Data Augmentation for Emotion Recognition Using a Conditional Wasserstein GAN. 2018 40th Annual International Conference of the IEEE Engineering in Medicine and Biology Society (EMBC), 2535–2538. \url{https://doi.org/10.1109/EMBC.2018.8512865}

M. Teplan. (2002). FUNDAMENTALS OF EEG MEASUREMENT. Institute of Measurement Science, Slovak Academy of Sciences, Dúbravská Cesta 9, 841 04 Bratislava, Slovakia, MEASUREMENT SCIENCE REVIEW, Volume 2, Section 2, 2002.

Minguillon, J., Lopez-Gordo, M. A., \& Pelayo, F. (2017). Trends in EEG-BCI for daily-life: Requirements for artifact removal. Biomedical Signal Processing and Control, 31, 407–418. \url{https://doi.org/10.1016/j.bspc.2016.09.005}

Neuper, C., Müller-Putz, G. R., Scherer, R., \& Pfurtscheller, G. (2006). Motor imagery and EEG-based control of spelling devices and neuroprostheses. In C. Neuper \& W. Klimesch (Eds.), Progress in Brain Research (Vol. 159, pp. 393–409). Elsevier. \url{https://doi.org/10.1016/S0079-6123(06)59025-9}

Nichol, A., \& Dhariwal, P. (2021). Improved Denoising Diffusion Probabilistic Models (arXiv:2102.09672). arXiv. \url{https://doi.org/10.48550/arXiv.2102.09672}

Niedermeyer, E., \& Silva, F. H. L. da. (2005). Electroencephalography: Basic Principles, Clinical Applications, and Related Fields. Lippincott Williams \& Wilkins.

Paluš, M. (1996). Nonlinearity in normal human EEG: Cycles, temporal asymmetry, nonstationarity and randomness, not chaos. Biological Cybernetics, 75(5), 389–396. \url{https://doi.org/10.1007/s004220050304}

Paszke, A., Gross, S., Chintala, S., Chanan, G., Yang, E., DeVito, Z., Lin, Z., Desmaison, A., Antiga, L., \& Lerer, A. (2017). Automatic differentiation in PyTorch. \url{https://openreview.net/forum?id=BJJsrmfCZ}

Penghai, L., \& Baikun, W. (2007). A Study on EEG Alpha Wave-based Brain-Computer Interface Remote Control System. 2007 International Conference on Mechatronics and Automation, 3179–3184. \url{https://doi.org/10.1109/ICMA.2007.4304070}

Prasad, T., \& Tvk, R. (2014). Survey on EEG Signal Processing Methods.

Riding, R. J., Glass, A., Butler, S. R., \& Pleydell‐Pearce, C. W. (1997). Cognitive Style and Individual Differences in EEG Alpha During Information Processing. Educational Psychology, 17(1–2), 219–234. \url{https://doi.org/10.1080/0144341970170117}

Sazgar, M., \& Young, M. G. (2019). EEG Artifacts. In M. Sazgar \& M. G. Young (Eds.), Absolute Epilepsy and EEG Rotation Review: Essentials for Trainees (pp. 149–162). Springer International Publishing. \url{https://doi.org/10.1007/978-3-030-03511-2_8}

Scherer, R., Muller, G. R., Neuper, C., Graimann, B., \& Pfurtscheller, G. (2004). An asynchronously controlled EEG-based virtual keyboard: Improvement of the spelling rate. IEEE Transactions on Biomedical Engineering, 51(6), 979–984. \url{https://doi.org/10.1109/TBME.2004.827062}

Shovon, T. H., Nazi, Z. A., Dash, S., \& Hossain, Md. F. (2019). Classification of Motor Imagery EEG Signals with multi-input Convolutional Neural Network by augmenting STFT. 2019 5th International Conference on Advances in Electrical Engineering (ICAEE), 398–403. \url{https://doi.org/10.1109/ICAEE48663.2019.8975578}

Smit, D. J. A., Boomsma, D. I., Schnack, H. G., Pol, H. E. H., \& Geus, E. J. C. de. (2012). Individual Differences in EEG Spectral Power Reflect Genetic Variance in Gray and White Matter Volumes. Twin Research and Human Genetics, 15(3), 384–392. \url{https://doi.org/10.1017/thg.2012.6}

Somepalli, G., Singla, V., Goldblum, M., Geiping, J., \& Goldstein, T. (2022). Diffusion Art or Digital Forgery? Investigating Data Replication in Diffusion Models (arXiv:2212.03860). arXiv. \url{https://doi.org/10.48550/arXiv.2212.03860}

Tallon-Baudry, C., \& Bertrand, O. (1999). Oscillatory gamma activity in humans and its role in object representation. Trends in Cognitive Sciences, 3(4), 151–162. \url{https://doi.org/10.1016/S1364-6613(99)01299-1}

Torma, S., \& Szegletes, D. L. (2023). Brain Signal Generation and Data Augmentation with a Single-Step Diffusion Probabilistic Model. \url{https://openreview.net/forum?id=woOQ5Hb1oOF}

Wang, F., Wu, S., Zhang, W., Xu, Z., Zhang, Y., Wu, C., \& Coleman, S. (2020). Emotion recognition with convolutional neural network and EEG-based EFDMs. Neuropsychologia, 146, 107506. \url{https://doi.org/10.1016/j.neuropsychologia.2020.107506}

Wang, J., Feng, Z., \& Lu, N. (2017). Feature extraction by common spatial pattern in frequency domain for motor imagery tasks classification. 2017 29th Chinese Control And Decision Conference (CCDC), 5883–5888. \url{https://doi.org/10.1109/CCDC.2017.7978220}

Wang, X.-W., Nie, D., \& Lu, B.-L. (2011). EEG-Based Emotion Recognition Using Frequency Domain Features and Support Vector Machines. In B.-L. Lu, L. Zhang, \& J. Kwok (Eds.), Neural Information Processing (pp. 734–743). Springer. \url{https://doi.org/10.1007/978-3-642-24955-6_87}
Wong, S. C., Gatt, A., Stamatescu, V., \& McDonnell, M. D. (2016). Understanding Data Augmentation for Classification: When to Warp? 2016 International Conference on Digital Image Computing: Techniques and Applications (DICTA), 1–6. \url{https://doi.org/10.1109/DICTA.2016.7797091}

\end{document}